# On-chip Acousto Thermal Shift Assay for Rapid and Sensitive Assessment of Protein Thermodynamic Stability

Yonghui Ding[1], Kerri A. Ball[2], Kristofor J. Webb[2], Yu Gao[1], Angelo D'Alessandro[3], William M. Old[2], Michael H.B. Stowell[2], Xiaoyun Ding[1,*]

[1] Paul M. Rady Department of Mechanical Engineering, University of Colorado, Boulder, CO 80309, USA

[2] Department of Molecular, Cellular and Developmental Biology, University of Colorado, Boulder, CO 80309, USA

[3] Department of Biochemistry and Molecular Genetics, University of Colorado Anschutz Medical Campus, Aurora, CO 80045, USA

* Corresponding Email: Xiaoyun.Ding@Colorado.edu

**Abstract:** Thermal shift assays (TSAs) have been extensively used to study thermodynamics of proteins and provide an efficient means to assess protein-ligand binding or protein-protein interaction. However, existing TSAs have limitations such as time consuming, labor intensive, or low sensitivity. Here we introduce a novel acousto thermal shift assay (ATSA), the first ultrasound enabled TSA, for real-time analysis of protein thermodynamic stability. It capitalizes the novel coupling of unique acoustic mechanisms to achieve protein unfolding, concentration, and measurement on a single microfluidic chip within minutes. Compared to conventional TSA methods, our ATSA technique enabled ultra-fast (at least 30 times faster), highly sensitive (7-34 folds higher), and label-free monitoring of protein-ligand interactions and protein stability. ATSA paves new avenues for protein analysis in biology, medicine and fast diagnosis.

Protein-ligand interactions are not only involved in almost every process in biological systems, but are also key events in the external modulation of protein function by drugs[1]. Protein thermal shift assays (TSAs) are a set of techniques to investigate protein-ligand interactions by detecting the changes in thermodynamic stability of the protein under varying conditions, including ligand binding[2,3]. Multiple TSA's methods have been implemented, however they have a variety of drawbacks, mass spectrum based TSAs are very expensive and labor/time intensive while other fluorescence based TSA approaches suffer from sensitivity issues, limiting their capability to detect small thermal shifts, particularly for low abundance or large, complex proteins[4,5]. Other limitations include time-consuming workflow, expensive equipment, technical experience, a lack of reliant labels such as antibodies, and/or demand for large amounts of purified proteins[4-6]. These limitations present a barrier to the broader applications of TSA in life science and medicine, particularly in the areas sensitive to efficiency and cost such as pharmaceutical industry and fast diagnosis.

Here we developed the first thermal shift assay enabled by acoustic mechanisms, acousto thermal shift assay (ATSA), where we employed surface acoustic waves (SAWs) to unfold proteins and concentrate the precipitated proteins on a microfluidic chip by capitalizing the novel coupling of acoustic heating and acoustic forces. When an acoustic field is imposed on a fluid, it will exert acoustic forces on suspended particles induced by acoustic scattering and also on fluid causing acoustic streaming due to viscous attenuation. Such acoustic forces have been employed to manipulate fluid, particles, and cells[7-12]. Acoustic heating, resulted from viscous attenuation of the acoustic energy into the fluid, was typically considered as a major hurdle for biomedical applications because the temperature rise is usually not welcome for the biological samples. By contrast, a well-controlled acoustic heating can be a valuable asset in driving chemical and biological reactions[13-15]. In our ATSA, we take advantage of the acoustic heating for fast and precisely controlled temperature ramping to unfold and precipitate proteins. Meanwhile, the acoustic force drives the assembly of precipitated proteins along the nodes and/or antinodes of the standing acoustic field, leading to significantly enhanced local concentration and thereby signal amplitude. The protein unfolding was monitored by measuring the gray intensity of precipitated proteins as a function of SAW time to analyze the thermal stability of proteins. The assay time is dramatically reduced to less than 2 minutes relative to tens of minutes or hours for conventional TSAs. We demonstrate the capability and superior sensitivity, i.e. up to 34-fold higher than conventional TSAs, of this new technique in detection of thermal shifts upon protein-ligand bindings and in diagnosis of mutational protein diseases, e.g. sickle cell disease (SCD).

**Results and Discussion**

**ATSA device design and characterization.** Specifically, two identical SAWs were generated by applying a AC (alternating current) signal to a pair of interdigital transducers (IDTs) deposited on the surface of a lithium niobate piezoelectric substrate and formed a standing SAW within a $1 \times 10$ mm$^2$ polydimethylsiloxane (PDMS) microchannel that was bonded on top of the substrate between these two IDTs (Fig. 1a-i and 1b). Under SAW actuation (19.6 MHz, 3 Watt), the temperature of a small-volume protein solution (less than 2 μL) in phosphate-buffered saline (PBS) within the microfluidic channel can be rapidly increased from 23°C to 80°C within 100 s. Most proteins rapidly precipitate and aggregate after their unfolding[3] (Fig. 1a-ii). Meanwhile, this standing SAW assembles and concentrates precipitated protein along the acoustic pressure nodes and/or antinodes (Fig. 1a-iii). The gray intensity ($I_m$) of the precipitated and assembled proteins was analyzed and plotted as a function of SAW time, giving rise to a sigmoidal melting curve (Figure 1a-iv). The melting time ($t_m$) was determined as the time point when half of proteins were unfolded, i.e. herein a 50% change of gray intensity occurs where $I_m = (I_{max} + I_{min})/2$. Using this simple analysis, the time shift $\Delta t_m$ between two samples can be measured at a given SAW power and analyzed to compare conditions or ligands that stabilize or destabilize proteins. Overall, this new technique provides a rapid, simple and efficient workflow for analysis of changes in apparent melting curves.

We first demonstrated the unfolding and concentration of a purified protein of human hemoglobin (Hb) under SAW actuation (Figure 1c). Hb is the most abundant protein in red blood cells (RBC) and its interactions with other molecules in blood are critical to its functions[16]. The protein solutions in the microfluidic channel were visually clear before SAW actuation. When SAW actuation (19.6 MHz, 3 Watt) was powered on, proteins were first unfolded and precipitated, resulting from the acoustic heating, and the precipitated proteins were quickly assembled along nearby pressure nodes and anti-nodes under acoustic forces to form concentrated protein microfibers (Fig. 1c). Similar SAW-driven protein unfolding and concentration were also observed with the mixed protein solution of human blood plasma (Fig. 1d). Together, we demonstrated that the SAW enabled rapid protein unfolding, precipitation, and concentration, which present an unprecedented potential for protein thermal stability analysis.

**ATSA for assessment of protein-ligand binding.** We next explored the utility of such SAW-driven protein unfolding and concentration for analysis of protein-ligand binding. We measured and plotted the gray intensity of the precipitated and concentrated proteins as a function of SAW time in ATSA, as shown in Fig. 2a. In parallel, we performed two representative conventional TSA methods, differential scanning fluorimetry (DSF) and bicinchoninic acid (BCA) assays, to analyze the melting temperature ($T_m$) and its shift ($\Delta T_m$) upon protein-ligand binding. DSF, the most popular TSA method, utilizes dye fluorescence as a measure of protein unfolding[17], while BCA quantifies the amount of remaining soluble proteins after thermal unfolding[18]. Two compounds, chloride (Pal) and oxaloacetic acid (OAA), were selected due to their well-known interactions with Hb and CS, respectively[19,20]. Under SAW actuation (19.6 MHz, 3 Watts), the typical sigmoidal melting curves of Hb and Hb-Pal complexes were obtained and revealed that the $t_m$ of Hb (46.85 s) was reduced by 3.52 s and 7.30 s in the presence of 0.62 mM and 1.24 mM Pal, indicating the destabilization of Hb by Pal (Fig. 2a). However, DSF-TSA could not successfully produce such typical sigmoidal melting curves for Hb and Hb-Pal complexes due to the inference from the intrinsic fluorescence of both Hb and Pal[21] (Fig. 2b), exemplifying the superiority of our ATSA method. Alternatively, we performed another conventional TSA method, BCA assay, to analyze the $T_m$ shift of Hb against Pal. The results showed that the $T_m$ of Hb (59.1 °C) was decreased by 0.7 °C and 1.1 °C in the presence of 0.62 mM and 1.24 mM Pal (Fig. 2c). Under the same SAW actuation, the $t_m$ of CS (36.23 s) was increased by 10.45 s and 15.77 s in the presence of 1 mM and 2 mM OAA, due to their known stabilizing effect on CS. By contrast, the $T_m$ shift of CS as detected in DSF assay was small and not statistically significant although the typical sigmoidal melting curves were obtainable (Fig. 2b). Likewise, the $T_m$ of CS (60.3 °C) was increased by only 2.0 °C and 2.9 °C in the presence of 1 mM and 2 mM OAA as detected in the BCA assay (Fig. 2c). In addition, the analysis of thermal stability of Hb and Hb-Pal complex with various concentrations showed that the detected $t_m$ and $\Delta t_m$ were not visibly sensitive to concentrations varying from 124 µM to 3.875 µM in current ATSA (Supplementary Fig. S1).

To make a direct comparison between our ATSA and conventional TSA methods, we first evaluated their sensitivity by defining the relative shift ($\frac{\Delta t_m}{t_{m0}}$ or $\frac{\Delta T_m}{T_{m0}}$) and sensitivity (tangential slope). The results showed that our ATSA method remarkably enhanced sensitivity in monitoring the protein-ligand binding compared to the conventional DSF-TSA and BCA-TSA methods (Fig. 2d). The sensitivity was dependent on the protein type and compound concentration. Specifically, our ATSA achieved a sensitivity of 7-fold higher in Hb-Pal binding and 9-fold higher in CS-OAA binding than BCA-TSA. More strikingly, our ATSA method produced a sensitivity of 34-fold higher than DSF-TSA method for CS-OAA binding. We further evaluated the precision as defined by the ratio between $\Delta t_m$ (or $\Delta T_m$) and standard deviation (s.d.) of $t_m$ (or $T_m$), i.e. $\frac{\Delta t_m}{s.d.}$ (or $\frac{\Delta T_m}{s.d.}$) in order (Fig. 2e). Our ATSA demonstrated much higher precision than DSF-TSA method while comparable to BCA-TSA method. Intriguingly, we found both sensitivity and precision were significantly compromised when the protein unfolding was induced under heating only condition instead of under SAW induced heating and concentration (Supplementary Fig. S2), suggesting the crucial role of acoustic concentration in ATSA.

In addition, we found that the magnitude of melting time shift $\Delta t_m$ could be facilely tuned by varying the SAW power in our ATSA (Fig. 3). The magnitude of $\Delta t_m$ between Hb and Hb-Pal complex were significantly increased and the $\Delta t_m$ became more viable by lowering the SAW power from 3 W to 2.5 W or 2 W (Fig. 3a, b), although the relative shifts seemed not to be visibly sensitive to SAW power (Fig. 3c). This tunability was attributed to the slower heating profiles under lower SAW powers (Fig. 3d). It would benefit the detection of marginal thermal shifts upon ligand bindings that might not be distinctly revealed by conventional TSA methods.

**Thermal shift assay for assessment of protein mutation.** These promising results encouraged us to further investigate the potential of our ATSA method for diagnosis of mutational protein diseases, which are associated with protein misfolding and subsequent thermal stability changes[22]. Sickle cell disease (SCD) is one example which affects millions worldwide and is caused by polymerization of sickle Hb in individual RBCs[23]. However, the lack of practical diagnostic approach leads to an inability to early treatment and high childhood mortality especially in resource-limited areas[24]. Thermodynamic instability of sickle Hb is characteristic of SCD[25,26]. We examined whether our technique could distinguish the Hb stability profiles with red blood cell (RBC) lysate from healthy (Ctrl) and SCD human donors (Fig. 4). The melting time $t_m$ of RBC lysates from healthy and SCD donors were 52.13 s and 45.98 s respectively, producing an apparent shift of $\Delta t_m$ in our ATSA method (Fig. 4a). By contrast, the melting curves produced by the conventional BCA-TSA method were almost overlapped and no significant melting temperature shift $\Delta T_m$ was detected between healthy and SCD RBC lysates, indicating its inefficacy to distinguish them (Fig. 4b). In addition, under SAW actuation, the parallelly aligned protein patterns were formed in the microfluidic channels and no visible difference in their morphology was observed between healthy and SCD Hb (Fig. 4c). Our technique demonstrates here a potential application as a new and promising point-of-care platform for rapid and highly sensitive diagnostic tool of SCD, although the further optimization is required for clinical use.

**Conclusion.** To conclude, the on-chip acousto thermal shift assay (ATSA technique reported here represents a novel thermal shift assay strategy and demonstrates unprecedented speed and sensitivity for label-free analysis of protein thermodynamic stability in real time. It requires less sample volume and is faster than any current TSA methods in measuring single protein melting curve without any needs of molecular markers, allowing its broad potential applications in fast diagnosis. The superior sensitivity to protein stability than conventional fluorescence based TSA methods shed light on its great potential in measuring quantitative and precise binding affinity, which will be one focus in our future work. Our next focus will also cover the fundamental study of acoustic effect on protein thermodynamic stability. Overall, with its high compatibility to automatic processing and smart phone, we envision that this novel ATSA system will profoundly benefit a plethora of applications in fundamental biomedical research, drug industry and fast diagnosis.

## Method

### Chemicals

Two proteins, i.e. human hemoglobin (Hb; Millipore Sigma, St. Louis, MO, USA) and citrate synthase (CS) from porcine heart (Millipore Sigma), and two corresponding compounds, i.e. palmatine chloride (Pal; Santa Cruz Biotechnology, Santa Cruz, CA, USA) and oxaloacetic acid (OAA; Millipore Sigma) were primarily used in this study. Human whole blood samples with K2 EDTA as anticoagulant were purchased from Zen-Bio Inc. (Research Triangle Park, NC, USA) and stored in 4 °C (always used within 3-14 days after collection). Human blood plasma was obtained by centrifuging the human whole blood samples at 500 $g$ for 10 min in 4 °C. Sickle cell diseased (SCD) red blood cell lysate were obtained from patients with SCD, upon receiving written informed consent and in conformity with the declaration of Helsinki under protocol approved by the Duke University Medical Center (no. NCT02731157) as described previously[27]. All abovementioned chemicals were dissolved or diluted using Dulbecco's Phosphate-Buffered Saline (DPBS; Thermal Fisher Scientific, Hampton, NH, USA) except Pal that was first dissolved in dimethyl sulfoxide (DMSO; Millipore Sigma) but then further diluted in DPBS. The DMSO concentration is 2% (vol/vol) in final Hb-Pal complex solution in order to prevent the potential damage to the protein by contact with high concentrations of DMSO. Fluorescent microspheres (polystyrene, 7.5 μm diameter) were obtained from Bangs Laboratories Inc. (Fishers, IN, USA).

### Device Fabrication

The SAW was generated and propagating on piezoelectric 128º Y-cut X-propagating lithium niobate (LiNbO$_3$) wafer (500 μm thick). The device consisted of a pair of interdigitated transducers (IDTs) in parallel in order to generate two series of identical SAWs propagating in the opposite direction, producing the standing SAW. Each IDT consists of 30 pairs of electrodes (Cr/Au, 5/100 nm) with the width of electrode finger of 50 μm, pitch of 100 μm, and an aperture of 10 mm, yielding a frequency of approximately 20 MHz for the propagating SAW. Although different IDTs were used in the project, the resonance frequencies of the IDTs are in the range between 19.5 and 19.6 MHz. A PDMS microchannel with height of 100 μm and width of 2 mm was then fabricated through a standard soft-lithography and model-replica procedure. Lastly, both the PDMS channel and the IDT substrate were treated with oxygen plasma and bonded together to form the final SAW device as shown in Fig. 1b.

**Acousto Thermal Shift Assay (ATSA) and Data Analysis**

The SAW device was mounted on the stage of an inverted microscope (ECLIPSE Ti-U, Nikon, Japan). A radio frequency (RF) signal was generated by a function generator (EXG Analog Signal Generator, Keysight, Santa Rosa, CA, USA) and amplified by an amplifier (403LA, Electronics & Innovation, Rochester, NY, USA). Five microliters of protein, plasma, red blood cell lysate or protein-compound mix solutions were injected into the channel before the RF signals were applied. A fast camera (ORCA-Flash4.0LT, Hamamatsu, Japan) was connected to the microscope to capture the process, and all the videos were recorded in 4 frames per second.

All image and videos processing were performed in ImageJ (National Institute of Health, Bethesda, MD, USA) in the same way as descried below. The same sized regions of interest (ROIs) were traced around the perimeter of each pattered protein fiber in order to monitor the gray intensity and its change during the course of protein melting and aggregation. At least five ROIs were selected and characterized for each video. Melting time was defined as the time point when there is 50% gray intensity change $I_m = (I_{max} + I_{min})/2$.

**Conventional Thermal Shift Assay**

Two conventional methods were adopted for thermal shift assay: SYPRO differential scanning fluorimetry (DSF) assay and bicinchoninic acid (BCA) assay.

SYPRO DSF assay: SYPRO Orange melting curves were collected using the 7900HT Fast Real-Time PCR System. The SYPRO Orange fluorescent signal is detectable using the calibration setting for the ROX filter. Melting curves were performed using 1 mg/mL of protein with a 1:2500 dilution of SYPRO Orange (Molecular Probes Inc #S-6651) in 100 mM PBS, pH 7.4, using a minimum of 4 replicates. A 1% ramp rate from 25°C to 95°C was utilized during data collection. Drug concentrations are as indicated. To analyze melting curves, the fluorescence was normalized to the starting temperature and to no protein controls. The data was then scaled to interval (0,1) and then the replicates were averaged, and standard deviation calculated.

BCA assay: For thermal gradient profiling a gradient program was created using a PTC-200 thermal cycler (MJ Research, Reno, NV, USA) to cover the temperature points indicated in each figure. A PCR plate was prepared with 25 μL per well of recombinant protein or lysate and sealed (4titude Random Access, PN 4ti-0960/RA 96-well plate). The plates were spun at 1200 *g* for two minutes at 4 °C, and then kept at 4 °C prior to use. The plate was placed in the thermal cycler with the heated lid closed for 3 minutes and was then spun at 1200 *g* for two minutes to remove any condensation. The PCR tubes were removed from the PCR plate, carefully placed in 1.5 mL tubes, and spun at 21,000 *g* for 30 min at 4 °C to pellet the aggregate protein. Supernatant was carefully removed from each tube and placed in a clean, low-retention, 1.5 mL tubes. 10 ul of solution was removed and a Pierce BCA protein assay kit (PN 23225) was used for the determination of the total protein in each sample.

**Statistical Analysis**

All data were expressed as means ± SD. The data (melting time and melting temperature) from multiple runs (n ≥ 3) were plotted using Graphpad Prism 8.0 software (GraphPad Software Inc., La Jolla, CA, USA). The BCA-TSA data were fitted using a Sigmoidal dose-response (variable slope) curve fit. Unpaired t-test or ordinary one-way ANOVA with Tukey's multiple comparison test was used to analyze statistical significance. A p-value < 0.05 was considered statistically significant. Within the figures, the significance was denoted by the following marks: * or # for $p < 0.05$; ** or ## for $p < 0.01$; and *** or ### $p < 0.001$.

**Acknowledgements**

The ATSA devices were fabricated in JILA clean room at University of Colorado Boulder. This work was supported by startup funding of University of Colorado Boulder.

**Author Contributions**

Y.D. and X.D. conceived and designed the research and wrote the manuscript. Y.D. performed the experiments. K.A.B., K.J.W., W.M.O., and M.H.B.S. designed and performed control experiments of DSF-TSA and BCA-TSA. A.D. provided SCD RBC lysate sample and related support. All authors contributed to the discussion of results and manuscript preparation.

**Competing interests**

The authors declare no competing financial interests.

# Figures

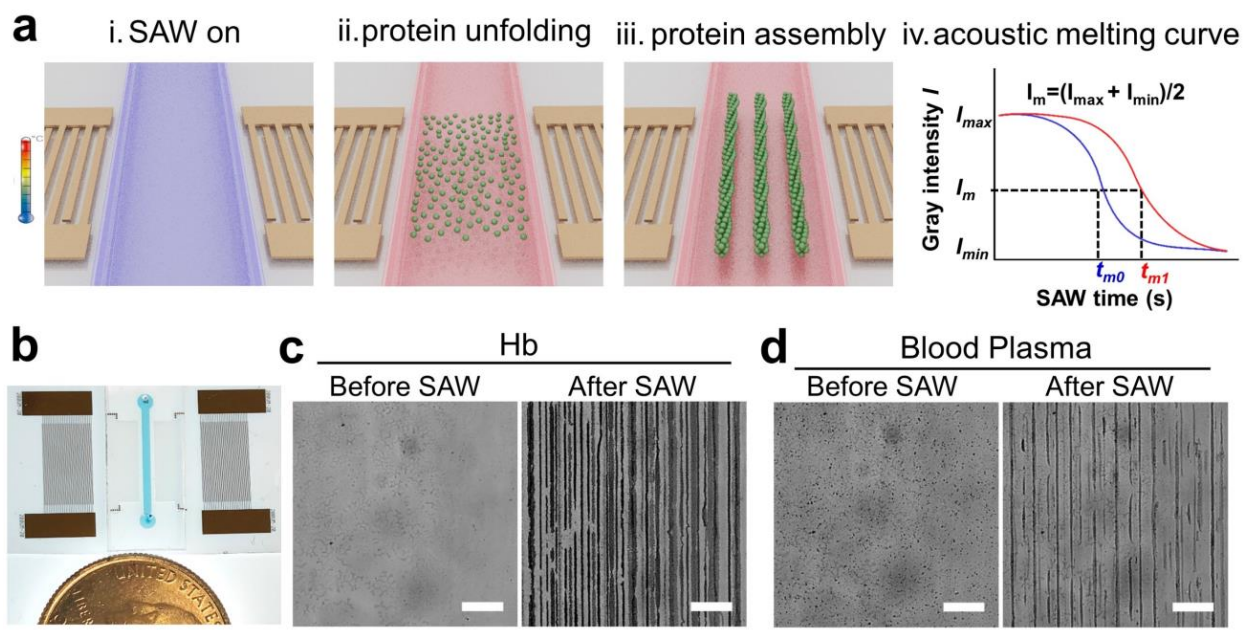

**Fig. 1. Working mechanism of acousto thermal shift assay (ATSA). a,** Schematic illustrating the working principle: (i) standing SAW formed in between two IDTs, (ii) protein unfolding and precipitation induced by acoustic heating within a microfluidic channel, (iii) precipitated proteins were assembled and concentrated along nodes and/or antinodes of standing acoustic field, and (iv) acoustic melting curves was generated by analyzing the gray intensity of precipitated and assembled protein as a function of SAW time, where the melting time ($t_m$) was determined as the time point at which the 50% change of gray intensity ($\Delta I$) occurs, i.e. $I_m = (I_{max} + I_{min})/2$. **b,** Photo of an ATSA device realized by bonding PDMS and lithium niobate wafer with a pair of IDTs. **c, d,** Acoustic-driven protein unfolding, precipitation, and assembly as demonstrated by optical images of purified protein, i.e. hemoglobin (Hb) in **c**, and mixed proteins, i.e. blood plasma in **d**, before and after SAW actuation. Scale bars: 200 μm.

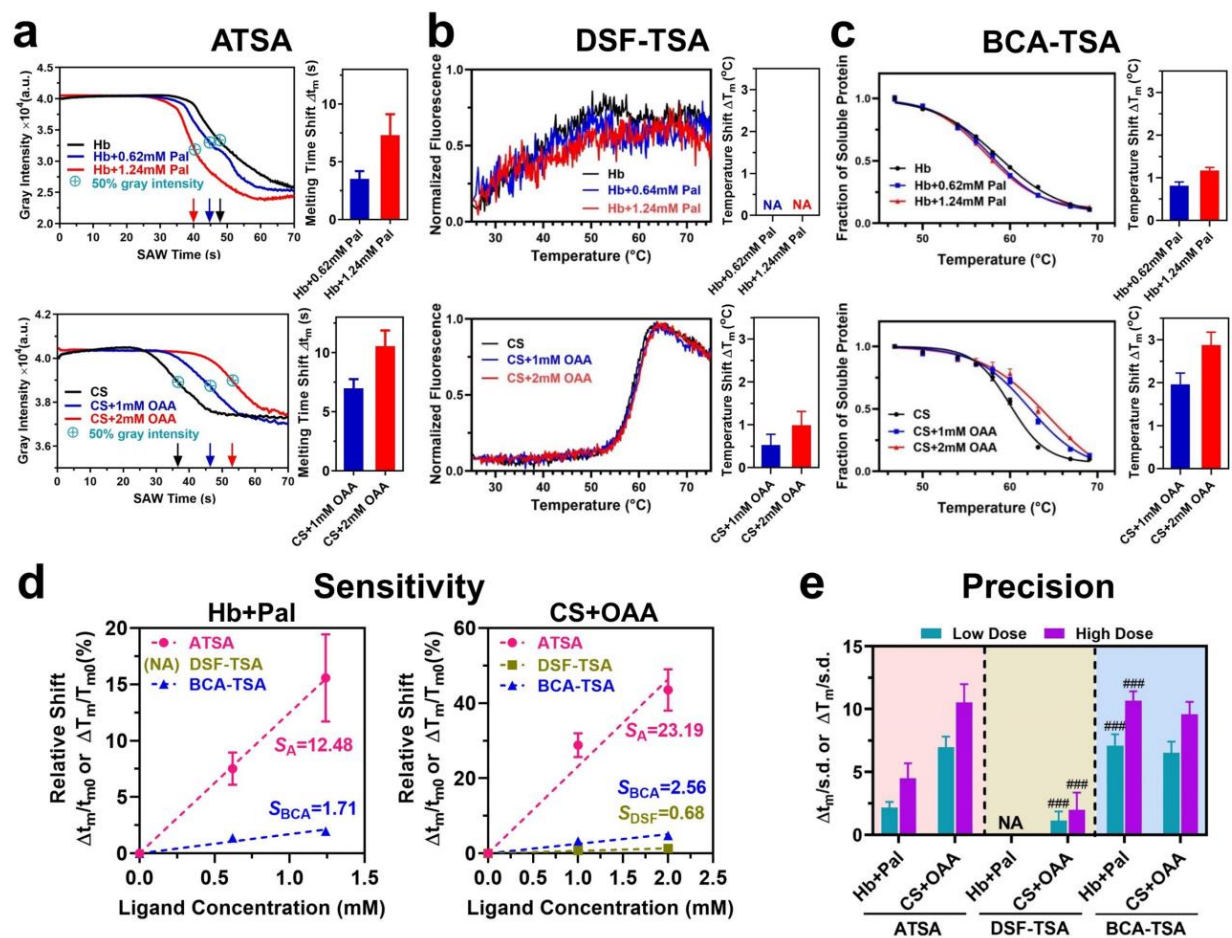

**Fig. 2. The ATSA enables rapid and sensitive assessment of protein-ligand binding and protein stability. a, b, c,** Representative melting curves and melting time shifts $\Delta t_m$ detected by our ATSA technique in **a** (n = 4), and melting temperature shifts $\Delta T_m$ detected by two conventional TSA methods, i.e. SYPRO differential scanning fluorimetry (DSF) assay in **b** (n = 4) and bicinchoninic acid (BCA) in **c** (n = 3), for two purified proteins, Hb and CS, in the absence or presence of their corresponding binding ligands, i.e. palmatine chloride (Pal) and oxaloacetic acid (OAA). The arrows in **a** indicate the melting time at which the 50% change of gray intensity ($I_m = (I_{max} + I_{min})/2$) occurs for each curve. The Hb concentration of 31 μM and the CS concentration of 15 μM were used in these tests. **d,** The relative shift ($\frac{\Delta t_m}{t_{m0}}$ or $\frac{\Delta T_m}{T_{m0}}$) of proteins upon ligand binding as a function of ligand concentration. The dashed lines represented linear regression curve fit of the data and their tangential slope was defined as sensitivity (mM$^{-1}$) of TSAs. **e,** Our ATSA

technique showed better or comparable precision, which is characterized by the fold differences between the detected shift and standard deviations (s.d.) of melting time or temperature, i.e. ($\frac{\Delta t_m}{\text{s.d.}}$ or $\frac{\Delta T_m}{\text{s.d.}}$), in analysis of thermal shifts than the conventional DSF-TSA or BCA-TSA methods. All error bars represent standard deviation (s.d.). In **b, d, e,** NA indicates the data of thermal shifts between Hb and Hb-Pal complexes is not available for DSF-TSA. In **e**, ### $p < 0.001$ *versus* ATSA method.

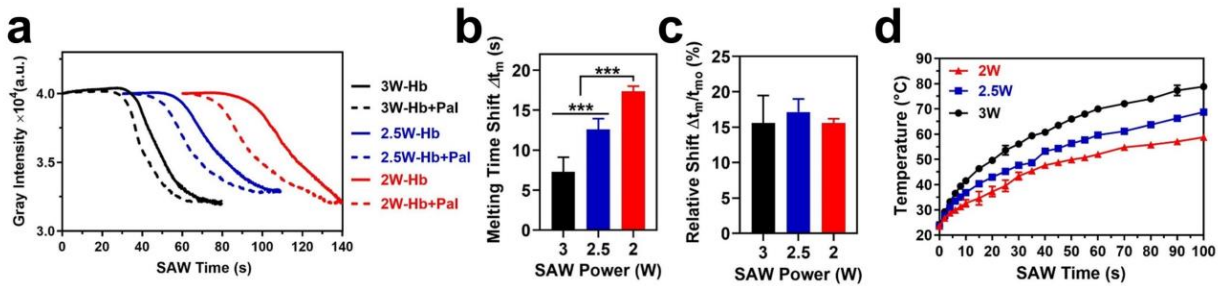

**Fig. 3. The magnitudes of melting time shift $\Delta t_m$ are tunable by adjusting SAW power in ATSA. a,** Representative melting curves and **b,** analysis of melting time shift $\Delta t_m$ (n = 4) of Hb and Hb-Pal complexes under various SAW power, 3 Watt, 2.5 Watt, and 2 Watt. The magnitude of melting time shift $\Delta t_m$ was remarkably increased by decreasing the SAW power. **c,** The relative shift seems not to be dependent on the SAW power. **d,** Acoustic heating effects are sensitive to the SAW power, i.e. lower SAW power resulted in slower temperature increase (n = 3), which is responsible for this tunability. All error bars represent standard deviation (s.d.). *** $p < 0.001$.

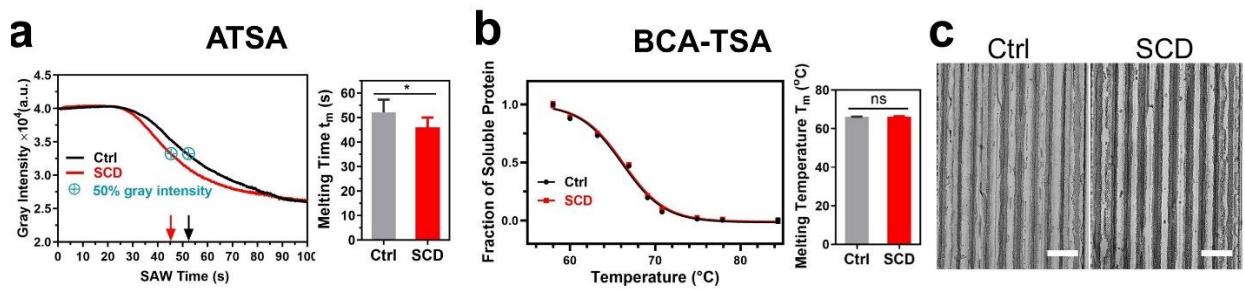

**Fig. 4. The ATSA allows sensitive detection of thermal shifts between healthy and sickled red blood cell (RBC) lysates, providing a new point-of-care platform for diagnosis of sickle cell disease (SCD). a, b,** Differences in protein stability between healthy and sickled RBC lysates (SCD) are detectable by our ATSA method while not detectable by a conventional TSA method, i.e. BCA assay, as shown by representative melting curves and analysis of melting time $t_m$ (n = 4) or melting temperature $T_m$ (n = 3). **c,** Optical images of patterned protein microfibers of healthy and sickled RBC lysates showed similar morphology. All error bars represent standard deviation (s.d.). In **a**, * $p < 0.05$. In **b,** ns means no significant difference ($p > 0.05$). Scale bars: 200 μm.

**References**

1. Frederick, K. K., *et al*. Conformational entropy in molecular recognition by proteins. *Nature* **448**, 325-329 (2007).

2. Huber, K. V. *et al*. Proteome-wide drug and metabolite interaction mapping by thermal-stability profiling. *Nat. methods*, **12,** 1055-1057 (2015).

3. Vedadi, M. *et al.* Chemical screening methods to identify ligands that promote protein stability, protein crystallization, and structure determination. *Proc. Natl. Acad. Sci. USA* **103**, 15835-15840 (2006).

4. Huynh, K., & Partch, C. L. Analysis of protein stability and ligand interactions by thermal shift assay. *Curr. Protoc. Protein Sci.* **79**, 28-9 (2015).

5. Jafari, R. *et al.* The cellular thermal shift assay for evaluating drug target interactions in cells. *Nat. Protoc.* **9**, 2100 (2014).

6. Dart, M. L. *et al.* Homogeneous Assay for Target Engagement Utilizing Bioluminescent Thermal Shift. *ACS Med. Chem. Lett.* **9**, 546-551 (2018).

7. Ding, X., *et al.* Surface acoustic wave microfluidics. *Lab on a Chip* **13**, 3626-3649 (2013).

8. Friend, J. & Yeo, L. Y. Microscale acoustofluidics: Microfluidics driven via acoustics and ultrasonics. *Reviews of Modern Physics* **83**, 647 (2011).

9. Ding, X., *et al.* On-chip manipulation of single microparticles, cells, and organisms using surface acoustic waves. *Proc. Natl. Acad. Sci. USA* **109**, 11105-11109 (2012).

10. Collins, D. J. et al. Two-dimensional single-cell patterning with one cell per well driven by surface acoustic waves. *Nat. commun.* **6**, 8686 (2015).

11. Franke, T., *et al.* Surface acoustic wave actuated cell sorting (SAWACS). Lab on a Chip **10** 789-794. (2010)

12	Frommelt, T., *et al.* Microfluidic mixing via acoustically driven chaotic advection. *Physical review letters* **100**. 034502. (2008)

13	Kulkarni, K., *et al.* Surface acoustic waves as an energy source for drop scale synthetic chemistry. *Lab on a Chip* **9**, 754-755 (2009).

14	Reboud, J. *et al.* Shaping acoustic fields as a toolset for microfluidic manipulations in diagnostic technologies.*Proc. Natl. Acad. Sci. USA* **109**, 15162-15167 (2012).

15	Shilton, R. J. et al. Rapid and controllable digital microfluidic heating by surface acoustic waves. *Adv. Funct. Mater.* **25**, 5895-5901 (2015).

16	Nagy, E. *et al.* Red cells, hemoglobin, heme, iron, and atherogenesis. *Atertio. Thromb. Vasc. Biol.* **30**, 1347-1353 (2010).

17	Matulis, D., *et al.* Thermodynamic stability of carbonic anhydrase: measurements of binding affinity and stoichiometry using ThermoFluor. J. *Biochemistry* **44**, 5258-5266 (2005).

18	Brown, R. E., *et al.* Protein measurement using bicinchoninic acid: elimination of interfering substances. *Anal. Biochem.* **180**, 136-139 (1989).

19	Liu, B., *et al.* Studies on the interaction of palmatine hydrochloride with bovine hemoglobin. *Luminescence* **29**, 211-218 (2014).

20	Niesen, F. H., *et al.* The use of differential scanning fluorimetry to detect ligand interactions that promote protein stability. *Nat. Protoc.* **2**, 2212 (2007).

21	Alpert, B., *et al.* Tryptophan emission from human hemoglobin and its isolated subunits. *Photochem. Photobiol.* **31**, 1-4 (1980).

22	Cohen, F. E. & Kelly, J. W. Therapeutic approaches to protein-misfolding diseases. *Nature* **426**, 905-909 (2003).

23	Piel, F. B., *et al.* Sickle cell disease. *New Engl. J. Med.* **376**, 1561-1573 (2017).

24	Ilyas, S., *et al.* Emerging Point-of-Care Technologies for Sickle Cell Disease Diagnostics. *Clin. Chim. Acta* (2019).

25	Tam, M. F. *et al.* Sickle cell hemoglobin with mutation at αHis-50 has improved solubility. *J. Biol. Chem.* **290**, 21762-21772 (2015).

26	Meng, F. *et al.* Substitutions in the β subunits of sickle-cell hemoglobin improve oxidative stability and increase the delay time of sickle-cell fiber formation. *J. Biol. Chem.* **294**, 4145-4159 (2019).

27	Culp-Hill, R. *et al.* Effects of red blood cell (RBC) transfusion on sickle cell disease recipient plasma and RBC metabolism. *Transfusion* **58**, 2797-2806 (2018)